\documentclass[11pt]{article}
\usepackage[utf8]{inputenc}
\usepackage{amsmath,amssymb,amsfonts}
\usepackage{graphicx}
\usepackage{authblk}
\usepackage{hyperref}
\usepackage{cite}
\usepackage{geometry}
\title{An Agent-Based Framework for Automated Higher-Voice Harmony Generation}

\author[1]{Nia D'Souza Ganapathy} 
\author[2]{Arul Selvamani Shaja, PhD}
\affil[1]{Bombay International School, Mumbai}
\affil[2]{Cisco Systems, Inc. San Jose, USA}

\date{\today}

\begin{document}

\maketitle

\begin{abstract}
The generation of musically coherent and aesthetically pleasing harmony remains a significant challenge in the field of algorithmic composition. This paper introduces an innovative Agentic AI-enabled Higher Harmony Music Generator, a multi-agent system designed to create harmony in a collaborative and modular fashion. Our framework comprises four specialized agents: a Music-Ingestion Agent for parsing and standardizing input musical scores; a Chord-Knowledge Agent, powered by a Chord-Former (Transformer model), to interpret and provide the constituent notes of complex chord symbols; a Harmony-Generation Agent, which utilizes a Harmony-GPT and a Rhythm-Net (RNN) to compose a melodically and rhythmically complementary harmony line; and an Audio-Production Agent that employs a GAN-based Symbolic-to-Audio Synthesizer to render the final symbolic output into high-fidelity audio. By delegating specific tasks to specialized agents, our system effectively mimics the collaborative process of human musicians. This modular, agent-based approach allows for robust data processing, deep theoretical understanding, creative composition, and realistic audio synthesis, culminating in a system capable of generating sophisticated and contextually appropriate higher-voice harmonies for given melodies.

\textbf{Keywords:} Multi-Agent Systems, Transformer Models, Generative Pre-trained Transformer (GPT), Symbolic Music Processing, Recurrent Neural Networks (RNN), Generative Adversarial Networks (GAN), Symbolic-to-Audio Synthesis
\end{abstract}

\section{Introduction}
Harmony is an integral component of music, creating a mixture of sounds to make compositions richer and more layered. Rooted in the principles of traditional music theory, harmonies often come to skilled composers intuitively. However, creating harmonies can be complex and time-consuming, especially for emerging musicians who may not have a strong grasp on its theoretical concepts. Thus, an automated harmony generation system would be a valuable tool, assisting composers in improving the quality of their compositions, as well as educating musicians in the process of creating harmonies.

Our proposed system takes music scores from the user and scans the melody and chord structures associated with it, producing harmonized versions of the score based on its knowledge of the theoretical ‘rules’ of harmony and traditional chordal frameworks. This allows it to accurately generate harmonies, bridging the gap between abstract music theory and practical application to compositions across genres.

This art of harmonizing a melody has remained a cornerstone of musical composition, woven into the tapestry of Western music to provide depth, color, and emotional resonance. From the foundational counterpoint of the Renaissance to the complex chromaticism of the Romantic era and the extended tonal language of jazz, it is a craft that transcends mere note selection. It involves the intricate dance of voice leading, the strategic management of tension and release, and the creation of a supportive yet independent musical line that enhances and elevates the primary melody. The creation of a compelling harmony is not simply a theoretical exercise but a deeply creative act, often considered a hallmark of compositional skill.

For centuries, this creative process has been the exclusive domain of human composers, who draw upon a blend of theoretical knowledge, stylistic convention, and innate musical intuition. However, the advent of artificial intelligence presents a fascinating new frontier: can we computationally model this nuanced and artistic process? Early attempts at algorithmic composition often relied on rigid, rule-based systems that, while capable of producing technically correct harmonies, frequently lacked the fluidity, context-awareness, and aesthetic appeal of human-composed music. These systems struggled to capture the subtle interplay between melodic contour, rhythmic impetus, and harmonic progression that is second nature to a skilled musician.

This paper addresses this challenge by moving beyond monolithic, rule-based approaches. We introduce a novel framework that reconceptualizes the task of harmony generation as a collaborative, agent-based process, mirroring the distinct roles within a human musical ensemble. Our work posits that by creating a system of specialized AI agents—each an expert in its own domain—we can achieve a more musically intelligent and artistically satisfying result. In our model, a "Librarian" agent first interprets the score, a "Theorist" agent deciphers the harmonic language, a "Composer" agent creatively crafts the harmony, and a "Conductor" agent renders the final performance.

This paper presents the design and implementation of an **Agentic AI-enabled Higher Harmony Music Generator**. We will detail the architecture of our multi-agent system and the sophisticated AI models that power each agent, including a Transformer-based **Chord-Former**, a generative **Harmony-GPT**, a recurrent **Rhythm-Net**, and a **Symbolic-to-Audio Synthesizer**. Through this modular approach, we aim to demonstrate a system that not only generates technically correct harmonies but also produces musically compelling results that are stylistically coherent and sensitive to the nuances of the input melody. The following sections will elaborate on the system's architecture, methodology, and experimental results, providing a comprehensive overview of its capabilities and potential as a tool for both musical research and creative application.

\section{Background and Related Work}
The automated generation of music, a subfield of both artificial intelligence and music technology, has a rich history. Research has progressed from deterministic, rule-based systems to highly sophisticated, data-driven deep learning models. This section provides an overview of the key paradigms and technological milestones that form the foundation for the proposed agentic harmony generator.

\subsection{Early Algorithmic and Rule-Based Approaches}
The earliest attempts at algorithmic composition were rooted in music theory and mathematics, employing rule-based systems, formal grammars, and stochastic processes. The ``Illiac Suite'' (1957) by Lejaren Hiller and Leonard Isaacson is a notable early example, which used a computer to generate musical scores based on probabilistic rules derived from species counterpoint and serialism \cite{hiller1959}. Another prominent approach involved Markov chains, which model the probability of transitioning from one musical event (like a note or chord) to the next. While capable of producing stylistically consistent short phrases, these models lack a conception of long-term structure, often resulting in compositions that meander without a sense of global coherence \cite{nierhaus2009}. These foundational methods, though limited, established the core principle of using formal systems to automate musical creation.

\subsection{The Rise of Recurrent Neural Networks for Sequential Modeling}
The advent of machine learning, particularly deep learning, marked a significant paradigm shift. Recurrent Neural Networks (RNNs) became a natural choice for music generation due to their inherent ability to model sequential data. By processing musical events one at a time and maintaining a hidden state, or ``memory,'' RNNs can learn the local patterns of melody and harmony from a corpus of existing music. The development of Long Short-Term Memory (LSTM) networks, a specialized type of RNN, was particularly crucial \cite{hochreiter1997}. LSTMs use a gating mechanism to mitigate the vanishing gradient problem, allowing them to learn longer-range dependencies than their vanilla RNN counterparts. Seminal work in this area demonstrated that LSTMs could successfully learn and generate melodies, blues improvisations, and even polyphonic music, capturing stylistic nuances with greater fidelity than previous stochastic models \cite{eck2002}. Despite this success, LSTMs still face challenges in maintaining musical coherence and structure over extended durations, such as the length of a full song. Our \textbf{Rhythm-Net} builds upon this legacy, using an RNN structure to specifically model the localized temporal relationships in rhythm.

\subsection{The Transformer Revolution in Symbolic Music}
The introduction of the Transformer architecture, with its self-attention mechanism, revolutionized the field of sequence modeling, including natural language processing and, subsequently, music generation \cite{vaswani2017}. Unlike RNNs, which process data sequentially, the self-attention mechanism allows a model to weigh the importance of all input tokens simultaneously, regardless of their position. This capability is exceptionally well-suited for capturing the complex, long-range dependencies that define musical form, such as motifs, phrases, and repeated sections. The Music Transformer was a landmark model that demonstrated the superiority of this approach, generating minute-long piano pieces with notable long-term coherence and structure \cite{huang2018}. Subsequent research has produced a variety of Transformer-based models for numerous musical tasks, including melody continuation, harmonization, and multi-instrumental arrangement. These models excel at understanding musical context, forming the architectural basis for our \textbf{Chord-Former} and \textbf{Harmony-GPT}.

\subsection{Generative Models for High-Fidelity Audio Synthesis}
While the aforementioned models operate in the symbolic domain (e.g., MIDI or MusicXML), a separate line of research has focused on the direct synthesis of audio waveforms. Generative Adversarial Networks (GANs) have proven to be a particularly effective architecture for this task. A GAN consists of two competing neural networks: a Generator, which creates new data samples, and a Discriminator, which tries to distinguish between real and generated samples. Through this adversarial process, the Generator learns to produce highly realistic outputs. In the audio domain, models like GANSynth have successfully used this framework to synthesize high-fidelity musical instrument sounds from symbolic inputs \cite{engel2019}. Unlike autoregressive models (e.g., WaveNet) that generate audio one at a time, GANs can often generate an entire sequence in parallel, making them significantly faster and more suitable for real-time applications. The \textbf{Symbolic-to-Audio Synthesizer} in our framework is a GAN-based model, designed to render the generated symbolic harmony into a realistic and expressive audio performance.

\subsection{The Agent-Based Paradigm in Computational Creativity}
While most research has focused on monolithic model architectures, our work adopts a multi-agent system (MAS) paradigm. A MAS is a decentralized system composed of multiple autonomous, interacting agents, each with specialized capabilities and goals \cite{wooldridge2009}. This approach offers modularity, scalability, and the ability to model complex, collaborative processes. In the context of computational creativity, a MAS can mimic the distinct roles found in a human creative team. For instance, agents can be specialized as performers, composers, or theorists, collaborating to produce a final artifact. Although the MAS paradigm is well-established in AI, its application to end-to-end music generation has been limited. Our framework is novel in its explicit use of specialized agents---the ``Librarian,'' ``Theorist,'' ``Composer,'' and ``Conductor''---to modularize the complex task of harmony generation, from score interpretation to final audio rendering.

\section{Methodology}
The core of our proposed system is a modular architecture that conceptualizes harmony generation not as a monolithic task, but as a collaborative process executed by a team of specialized, autonomous agents. This approach is grounded in the principles of Multi-Agent Systems (MAS), a paradigm in which decentralized agents interact to solve problems that are beyond their individual capabilities \cite{wooldridge2009}. By assigning distinct roles—parsing, theoretical analysis, creative composition, and audio rendering—to different agents, our framework mimics the functional specialization seen in human musical ensembles. This modularity enhances the system's robustness and allows for the integration of highly specialized AI models at each stage of the creative pipeline. The following sections will detail this multi-agent framework and the specific functions and underlying models of each constituent agent.

\subsection{System Architecture: A Multi-Agent Framework}
\begin{figure}[htbp]
    \centering
    \includegraphics[width=1.05\linewidth]{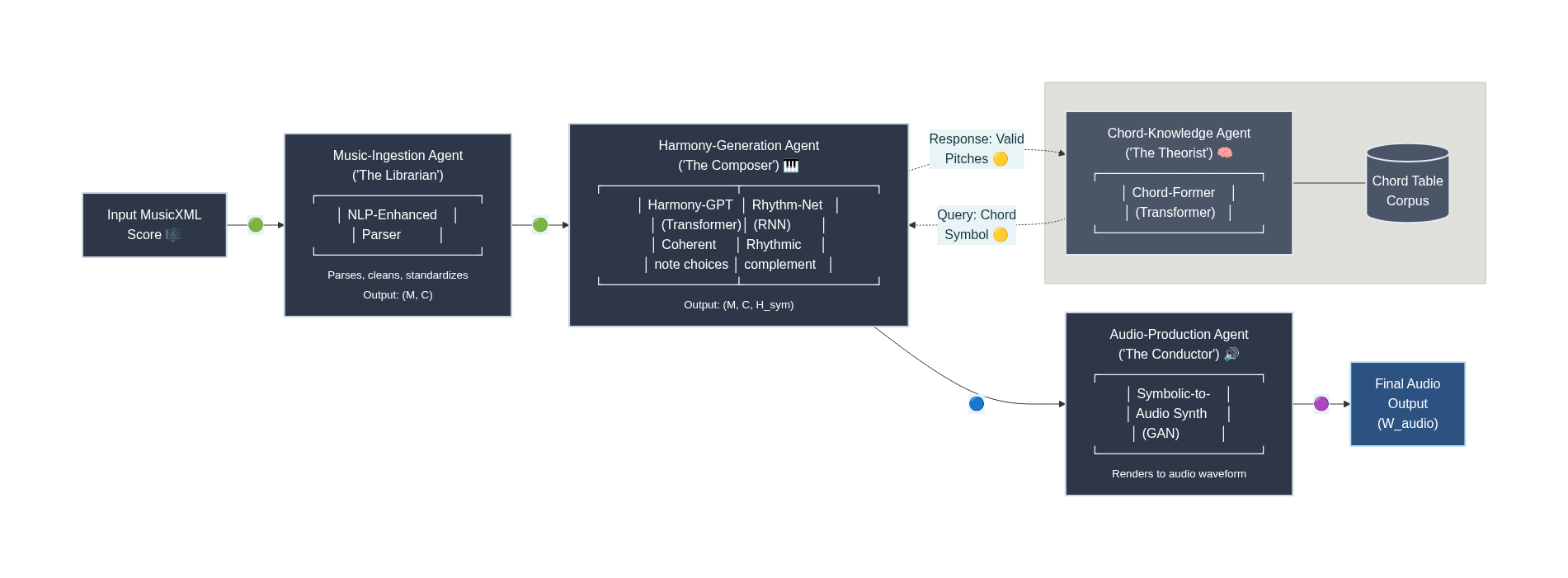}
    \caption{Agentic AI pipeline for music harmonization: MusicXML ingestion, knowledge-guided composition, and GAN-based audio synthesis.}
    \label{fig:A Multi-Agent Framework}
\end{figure}

Our system is designed as a sequential pipeline of four distinct agents: the Music-Ingestion Agent (the "Librarian"), the Chord-Knowledge Agent (the "Theorist"), the Harmony-Generation Agent (the "Composer"), and the Audio-Production Agent (the "Conductor"). The data, in the form of a musical piece, is passed from one agent to the next, with each agent performing a specific transformation that enriches and refines the musical content. The collaborative nature of this system allows for a sophisticated workflow, from initial symbolic representation to final audio synthesis, a concept that has shown promise in various computational creativity domains \cite{dahlstedt2015}.  The full pipeline is shown in Figure~\ref{fig:A Multi-Agent Framework}

We can formally represent the entire harmony generation process, $\mathcal{G}$, as a composition of functions, where each function corresponds to the transformation applied by an agent. Let the initial input be a musical score in MusicXML format, denoted as $S_{XML}$. The final output is an audio waveform, $W_{audio}$. The process can be defined as follows:

The overall system is a function $\mathcal{G}: S_{XML} \rightarrow W_{audio}$, which can be decomposed into the sequential operations of the agents:

\begin{equation}
    \mathcal{G} = A_{audio} \circ A_{harmony} \circ A_{ingest}
\end{equation}

Where:
\begin{itemize}
    \item $A_{ingest}$ is the function of the Music-Ingestion Agent, which parses the input score $S_{XML}$ and transforms it into a standardized, machine-readable format, $S_{std}$. This standardized format separates the melody line, $M$, from the sequence of chord symbols, $C$.
    \begin{equation}
        S_{std} = (M, C) = A_{ingest}(S_{XML})
    \end{equation}
    Here, $M = \{m_1, m_2, \dots, m_n\}$ is the sequence of melody notes, and $C = \{c_1, c_2, \dots, c_k\}$ is the sequence of chord symbols.

    \item $A_{harmony}$ represents the Harmony-Generation Agent. This agent takes the standardized data $S_{std}$ as input and generates a new symbolic score, $S_{harm}$, which includes the newly composed higher harmony line, $H_{sym}$. This agent internally consults the Chord-Knowledge Agent to determine the set of valid notes for each chord symbol.
    \begin{equation}
        S_{harm} = A_{harmony}(S_{std}) = (M, C, H_{sym})
    \end{equation}

    \item $A_{audio}$ is the function of the Audio-Production Agent, which takes the final symbolic score $S_{harm}$ and synthesizes it into the final audio waveform, $W_{audio}$.
    \begin{equation}
        W_{audio} = A_{audio}(S_{harm})
    \end{equation}
\end{itemize}

This agent-based architecture, illustrated in Figure 1, provides a clear and organized framework for the complex task of harmony generation. Each agent functions as an expert module, and their collective operation results in the final, harmonized musical piece. The subsequent sections will provide a detailed description of the internal workings and AI models of each agent.

\subsection{Data Corpus and Preprocessing}
The training of our system's models relies on two distinct types of musical data corpora. The first is a structured knowledge base for harmonic theory, while the second is a large-scale, diverse collection of musical scores for learning sequential and stylistic patterns.

\subsubsection{Chord Knowledge Corpus}
The foundation for the Chord-Knowledge Agent is a meticulously curated dataset. This corpus functions as a comprehensive dictionary, mapping a wide array of chord symbols directly to their constituent musical notes. It contains several hundred entries, encompassing standard triads (major, minor), seventh chords, extended harmonies (9ths, 11ths, 13ths), and various alterations (e.g., $\flat$5, $\sharp$9).

\textbf{Preprocessing} of this corpus is straightforward. The data is parsed into key-value pairs, where the chord symbol (e.g., `Xm7b5`) serves as the key and the corresponding note intervals (e.g., Root, Minor Third, Diminished Fifth, Minor Seventh) serve as the value. A vocabulary is constructed from all unique chord symbols, and each symbol is tokenized, or mapped to a unique integer index. This tokenized representation is used to train the \textbf{Chord-Former}, which learns the syntactic and semantic relationships between the symbols and their musical meanings.

\textbf{Preprocessing} of this large-scale corpus is performed by the Music-Ingestion Agent and involves several critical steps:
\begin{enumerate}
    \item \textbf{Data Filtering:} The dataset is scanned to select scores that are suitable for our task. We filter for pieces that contain both a clear melodic line and explicit chord annotations, discarding purely percussive or monophonic tracks.
    \item \textbf{Feature Extraction:} For each selected piece, we extract melody-chord pairs. The melody is extracted as a sequence of notes, while the chord progression is extracted as a corresponding sequence of chord symbols aligned with the melody.
    \item \textbf{Data Augmentation:} To increase the robustness and generalization of our models, we augment the dataset by transposing each piece into all twelve chromatic keys. This artificially expands the training data by a factor of 12 and helps the models learn harmonic relationships independent of a specific key.
\end{enumerate}

\subsection{Musical Data Representation}
To make symbolic music intelligible to our neural network models, we must convert it into a quantitative, numerical format. Our system employs a multi-faceted event-based representation that encodes pitch, rhythm, and harmony into a sequence of vectors. An event in our sequence represents a single musical note and its associated harmonic context.

Let a musical piece be represented as a sequence of $n$ events, $E = (e_1, e_2, \dots, e_n)$. Each event $e_i$ is a vector that encapsulates the core musical information:

\begin{equation}
    e_i = (\text{pitch}_i, \text{duration}_i, \text{onset}_i, \text{chord}_i)
\end{equation}

The components of this vector are represented as follows:

\begin{itemize}
    \item \textbf{Pitch:} Note pitches are represented using the standard MIDI note numbering scheme, an integer value from 0 to 127. A special value (e.g., 128) is reserved to denote a "rest" event.

    \item \textbf{Rhythm (Duration and Onset):} Rhythmic information is captured by two values. The \textbf{duration} of a note is quantized into discrete steps relative to a whole note (e.g., a value of 1.0 for a whole note, 0.25 for a quarter note). The \textbf{onset} represents the precise time, in beats, when the note occurs, measured from the beginning of the piece.

    \item \textbf{Harmony (Chord):} The harmonic context is represented by the chord symbol active at the onset of the note. As described in the preprocessing stage, each unique chord symbol in our corpus is mapped to an integer token via a vocabulary lookup. For a vocabulary $\mathcal{V}_C$, the chord symbol $c$ is represented as:
    \begin{equation}
        \text{chord}_i = \mathcal{V}_C(c)
    \end{equation}
\end{itemize}

This event-based sequence serves as the direct input for the Harmony-Generation Agent, providing the \textbf{Harmony-GPT} and \textbf{Rhythm-Net} with a rich, multi-dimensional understanding of the musical context from which to generate a new harmony line.

\subsection{The Agentic Workflow}
The four agents in our system operate in a sequential pipeline, creating a structured and hierarchical workflow. The musical data is progressively refined as it passes from the ingestion agent to the final audio production agent. This section details the specific role, underlying AI models, and formal operations of each agent within this collaborative process.

\subsubsection{Music-Ingestion Agent (The ``Librarian'')}
The workflow begins with the Music-Ingestion Agent, responsible for parsing and structuring the input MusicXML file. Its primary function is to act as a robust interpreter of symbolic notation, preparing the data for the downstream analytical and generative tasks. To handle the syntactic variations and potential ambiguities in MusicXML files, this agent is enhanced with an NLP-based parsing model trained to identify and standardize musical elements \cite{castellanos2020}.

Let the input MusicXML file be $S_{XML}$. The agent's transformation, $A_{ingest}$, can be formalized as:
\begin{equation}
    A_{ingest}: S_{XML} \rightarrow (M, C)
\end{equation}
Where $M = (m_1, m_2, \dots, m_n)$ is the extracted melody sequence, with each element $m_t$ being an event vector as defined in Section 3.3. Similarly, $C = (c_1, c_2, \dots, c_k)$ is the sequence of chord symbols, temporally aligned with the melody. The agent ensures that the data is clean, standardized, and ready for processing by the subsequent agents.

\subsubsection{Chord-Knowledge Agent (The ``Theorist'')}
The Theorist agent serves as the system's repository of harmonic knowledge. It is powered by the \textbf{Chord-Former}, a Transformer model trained on our Chord-Knowledge corpus. Its sole function is to provide the set of valid pitches for any given chord symbol. When queried by the Harmony-Generation Agent, the Theorist provides the raw material—the correct notes—from which a harmony can be constructed.

We can define the agent's function, $\mathcal{T}$, as a mapping from a chord symbol token, $c_j \in C$, to a set of valid MIDI pitch values, $P_j$:
\begin{equation}
    P_j = \mathcal{T}(c_j)
\end{equation}
For example, for the input symbol `Cmaj7`, the agent would return the set of pitches corresponding to C, E, G, and B. The Transformer architecture allows this agent to infer relationships between chords, enabling it to potentially derive the constituent tones for chord symbols not explicitly present in its training data, thus demonstrating a degree of music-theoretical generalization \cite{vaswani2017}.

\subsubsection{Harmony-Generation Agent (The ``Composer'')}
This is the creative nexus of the system. The Composer agent synthesizes the information from the Librarian and the Theorist to generate a novel  higher harmony line, $H_{sym}$. This agent's decision-making process is governed by two deep learning models: the \textbf{Harmony-GPT} for melodic and harmonic coherence, and the \textbf{Rhythm-Net} (RNN) for rhythmic complementarity.

The generation of the harmony sequence $H_{sym} = (h_1, h_2, \dots, h_n)$ is modeled as an autoregressive process. The agent aims to select the optimal harmony note $h_t$ at each time step $t$, conditioned on the melody sequence $M$, the chord sequence $C$, and the previously generated harmony notes $H_{<t}$. The selection is based on the probability distribution learned by the Harmony-GPT:
\begin{equation}
    p(h_t | M, C, H_{<t})
\end{equation}
The choice of note is constrained to the set of valid pitches $P_j$ provided by the Theorist for the chord $c_j$ active at time $t$. Simultaneously, the Rhythm-Net analyzes the rhythmic properties of $m_t$ to influence the duration and onset of the chosen harmony note $h_t$, ensuring the final harmony is not only melodically consonant but also rhythmically engaging.

\subsubsection{Audio-Production Agent (The ``Conductor'')}
The final stage is managed by the Conductor agent, which translates the complete symbolic score into a high-fidelity audio performance. This agent employs a \textbf{Symbolic-to-Audio Synthesizer}, a GAN-based model designed for realistic music synthesis \cite{engel2019}.

The agent receives the full symbolic representation, $S_{harm} = (M, C, H_{sym})$. The Generator network, $G$, of the GAN is tasked with mapping this symbolic data to an audio waveform, $W_{audio}$:
\begin{equation}
    W_{audio} = G(S_{harm}; \theta_g)
\end{equation}
where $\theta_g$ are the parameters of the Generator. A corresponding Discriminator network, $D$, is trained to distinguish between the synthesized audio from $G$ and real audio recordings from a target dataset. This adversarial training process compels the Generator to produce audio that is rich in timbre, dynamics, and articulation, effectively rendering a nuanced and expressive performance of the generated composition.

\subsection{Core AI Model Architectures}
The efficacy of our agent-based system is contingent upon the sophisticated deep learning models that empower each agent. These architectures are selected based on their proven strengths in handling specific types of data and tasks, from sequence-to-sequence translation to generative modeling. This section provides a detailed overview of the core AI models employed within our framework.

\subsubsection{Chord-Former: A Transformer for Harmonic Context}
The Chord-Knowledge Agent is powered by Chord-Former, a model based on the Transformer architecture \cite{vaswani2017}. Specifically, we employ an encoder-only structure, as the task is one of interpretation rather than generation of a new sequence. The input to the model is a sequence of characters representing a chord symbol (e.g., `G`, `m`, `7`, `b`, `9`). The model's objective is to map this symbolic input to a multi-hot encoded vector representing the set of constituent pitches in a 12-tone chromatic scale.

The core of the Chord-Former is the self-attention mechanism, which allows the model to weigh the importance of different characters within the chord symbol. For example, in `Xm7b5`, the `m` modifies the third, the `7` adds a minor seventh, and the `b5` alters the fifth. Self-attention enables the model to learn these complex, position-independent relationships, which are often challenging for sequential models like RNNs. The model is trained on the chord knowledge corpus, learning a robust mapping from symbolic notation to harmonic content.

\subsubsection{Harmony-GPT: Generative Melodic Harmonization}
The creative engine of the Harmony-Generation Agent is Harmony-GPT, a decoder-only Generative Pre-trained Transformer. This model is designed to function as a powerful sequence generator, leveraging the self-attention mechanism to capture long-range dependencies and complex patterns within musical compositions.

The model operates autoregressively, generating the harmony note $h_t$ at each time step $t$ by conditioning on a concatenated sequence of the input melody $M$ and the previously generated harmony notes $H_{<t}$. The model outputs a probability distribution over the entire vocabulary of possible notes. The probability of a harmony sequence $H_{sym}$ is thus the product of the conditional probabilities of its constituent notes:
\begin{equation}
    P(H_{sym} | M, C) = \prod_{t=1}^{n} p(h_t | M, C, H_{<t}; \theta)
\end{equation}
where $\theta$ represents the learned parameters of the Harmony-GPT model. During inference, a sampling strategy (such as top-k sampling or nucleus sampling) is used to select a note from this distribution, allowing for a balance between coherence and creativity \cite{holtzman2019}.

\subsubsection{Rhythm-Net: An RNN for Rhythmic Coherence}
To ensure that the generated harmony is rhythmically complementary to the melody, we employ Rhythm-Net, a Recurrent Neural Network (RNN), specifically a Long Short-Term Memory (LSTM) network \cite{hochreiter1997}. While Transformers excel at capturing global structure, RNNs are highly effective at modeling local, step-by-step temporal dependencies, making them ideal for the nuances of rhythmic interaction.

Rhythm-Net processes the sequence of melodic events in parallel with the Harmony-GPT. At each time step $t$, it takes the rhythmic information of the melody note $m_t$ (duration and onset) as input. Its function is to predict a suitable duration for the corresponding harmony note $h_t$. By maintaining a hidden state that carries information about the recent rhythmic context, the Rhythm-Net can learn conventional rhythmic patterns, such as syncopation or following the melodic rhythm, adding a crucial layer of musical sophistication to the generated output.

\subsubsection{Symbolic-to-Audio Synthesizer: A GAN for Audio Rendering}
The final translation from a symbolic score to a realistic audio waveform is performed by a Generative Adversarial Network (GAN) \cite{goodfellow2014}. Our Symbolic-to-Audio Synthesizer consists of two main components: a Generator ($G$) and a Discriminator ($D$).

The Generator, $G$, is a deep convolutional neural network designed to take the symbolic representation of the final harmonized piece, $S_{harm}$, and upsample it into a high-resolution audio waveform, $W_{audio}$. The Discriminator, $D$, is also a convolutional network, trained to distinguish between real audio samples from a training dataset of musical performances and the fake samples generated by $G$.

The two networks are trained in a minimax game, where $G$ aims to fool $D$, and $D$ aims to correctly classify the samples. The objective function for this adversarial training is:
\begin{equation}
    \min_{G} \max_{D} V(D, G) = \mathbb{E}_{x \sim p_{data}(x)}[\log D(x)] + \mathbb{E}_{z \sim p_{z}(z)}[\log(1 - D(G(z)))]
\end{equation}
where $x$ is a real audio sample and $z$ is the input symbolic data. This adversarial process forces the Generator to produce waveforms that are not just technically correct but also capture the timbral richness and subtle acoustic characteristics of real musical instruments.

\subsection{Training and Implementation Details}
The implementation of our system and the training of its constituent models were conducted using modern deep learning frameworks and hardware. This section outlines the technical specifications, training protocols, and optimization strategies employed.

\subsubsection{Technical Environment}
All models were implemented using the \textbf{PyTorch} deep learning framework \cite{paszke2019}, chosen for its flexibility in defining custom architectures and its robust support for GPU acceleration. The MusicXML parsing and manipulation were handled by the \textbf{music21} Python library \cite{cuthbert2010}. The training was executed on a high-performance computing cluster equipped with NVIDIA A100 GPUs, which significantly accelerated the training of the large Transformer and GAN models.

\subsubsection{Model Training and Optimization}
Each of the core AI models underwent a tailored training process to optimize its performance for its specific task.

\paragraph{Chord-Former:}
The Chord-Former was trained on the tokenized chord knowledge corpus. Due to the relatively small size of this dataset, we employed a simple yet effective training strategy. The model was trained using a cross-entropy loss function to predict the multi-hot encoded vector of chord tones from the input chord symbol. We used the Adam optimizer \cite{kingma2014} with a learning rate of $1 \times 10^{-4}$ and a batch size of 64. Training was conducted for 200 epochs, which was sufficient for the model to achieve near-perfect accuracy on a held-out test set of chord symbols.

\paragraph{Harmony-GPT:}
The Harmony-GPT, being the largest model, required the most extensive training. The model was first pre-trained on the general task of predicting the next musical event in the sequence, encompassing melody and harmony. It was then fine-tuned specifically on the task of predicting a harmony note, given the melody and chord context. The model was trained using a masked cross-entropy loss, optimizing only for the prediction of the harmony notes. We used the AdamW optimizer with a learning rate schedule that included a linear warm-up followed by a cosine decay. The training was distributed across four GPUs for a total of 200,000 steps.

\paragraph{Rhythm-Net:}
The Rhythm-Net LSTM was trained on the same dataset as the Harmony-GPT. Its objective was to predict the duration of a harmony note given the duration and onset of the corresponding melody note. We formulated this as a regression problem, using a Mean Squared Error (MSE) loss function to minimize the difference between the predicted and actual durations. The model was trained using the RMSprop optimizer with a learning rate of $1 \times 10^{-3}$. This model converged quickly due to its smaller size and the focused nature of its task.

\paragraph{Symbolic-to-Audio Synthesizer:}
Training the GAN-based synthesizer involved a specialized audio dataset. We used the NSynth dataset \cite{engel2017}, which contains a large corpus of high-quality, single-note audio samples from a wide range of musical instruments. The GAN was trained to generate audio waveforms corresponding to the pitch, duration, and instrument timbre specified in the symbolic score. The training followed the standard minimax procedure, using the Wasserstein GAN with Gradient Penalty (WGAN-GP) loss function for improved stability \cite{gulrajani2017}. The Generator and Discriminator were trained alternately using the Adam optimizer with distinct learning rates ($1 \times 10^{-4}$ for the Generator and $4 \times 10^{-4}$ for the Discriminator) to maintain a healthy adversarial dynamic.

\subsection{Advantages of the Agentic Framework}
The proposed multi-agent architecture for harmony generation offers several significant advantages over traditional, monolithic approaches to algorithmic composition. These benefits stem directly from its modular, decentralized, and functionally specialized design.

\begin{itemize}
    \item \textbf{Modularity and Extensibility:} By decomposing the complex problem of harmonization into a series of discrete sub-tasks, the framework becomes inherently modular. Each agent is a self-contained unit with a specific responsibility. This design greatly simplifies development, testing, and maintenance. Furthermore, the system is highly extensible. For instance, the Symbolic-to-Audio Synthesizer could be replaced with an improved model without affecting the other agents. Similarly, new agents, such as a Bass-Line Generation Agent or a Percussion Agent, could be seamlessly integrated into the pipeline to expand the system's capabilities into a full-fledged automated arranger.

    \item \textbf{Specialization and Expertise:} The agent-based approach allows for the deployment of highly specialized AI models precisely where they are most effective. Instead of relying on a single, general-purpose model, our system leverages the unique strengths of different architectures: the contextual understanding of Transformers for harmonic theory (Chord-Former, Harmony-GPT), the sequential prowess of RNNs for localized rhythm (Rhythm-Net), and the generative fidelity of GANs for audio synthesis. This functional specialization ensures that each aspect of the musical creation process is handled by a model optimized for that specific task, leading to a more musically nuanced and technically proficient final output.

    \item \textbf{Interpretability and Debugging:} The sequential nature of the pipeline provides clear points for inspection, enhancing the system's interpretability—a key aspect of Explainable AI (XAI) \cite{arrieta2020}. One can analyze the output of each agent individually: the standardized score from the "Librarian," the chord-tone sets from the "Theorist," and the symbolic harmony from the "Composer." This transparency makes it significantly easier to diagnose errors, understand the model's decision-making process, and fine-tune specific components of the creative workflow. This is a marked improvement over "black box" end-to-end models where the internal logic is often opaque.

    \item \textbf{Biomimicry of Human Collaboration:} The framework's design is inspired by and mirrors the collaborative process of human musicians. In a band or orchestra, different musicians have specialized roles (e.g., rhythm, harmony, melody), and the final piece is a product of their collective interaction. Our system operationalizes this concept by creating a virtual ensemble of AI agents. This biomimetic approach not only provides a logical and effective structure for the computational task but also aligns the system's creative process more closely with the proven methods of human artistry, a principle that has shown great promise in computational creativity \cite{saunders2019}.
\end{itemize}

These advantages position the agentic framework as a robust, flexible, and powerful paradigm for addressing the challenges of automated music composition, paving the way for more sophisticated and musically intelligent generative systems.

\section{Challenges and Limitations}
Despite the promising capabilities of the proposed agentic framework, it is essential to acknowledge its inherent challenges and limitations. These issues represent significant areas for future research and development within the field of algorithmic composition.

\begin{itemize}
    \item \textbf{Data Dependency and Stylistic Bias:} The creative output of our system is fundamentally constrained by the data on which its models are trained. Both the harmonic vocabulary of the Chord-Former and the compositional patterns learned by Harmony-GPT are direct reflections of their respective training corpora. So it may contain biases towards certain genres or styles. Consequently, the system may struggle to generate harmonies in styles that are underrepresented in the training data, potentially limiting its stylistic versatility and reinforcing existing musical conventions.

    \item \textbf{Computational Expense:} The use of state-of-the-art deep learning models, particularly large-scale Transformers and Generative Adversarial Networks, entails a significant computational cost. The training phases for Harmony-GPT and the Symbolic-to-Audio Synthesizer require substantial GPU resources and time. Furthermore, the inference or generation phase, especially for the audio synthesis, can be resource-intensive, making real-time application of the full pipeline a considerable technical challenge that may not be feasible on standard consumer hardware.

    \item \textbf{The Subjectivity of Musical Evaluation:} A persistent challenge in generative art is the difficulty of objective evaluation. While we can use quantitative metrics like perplexity or loss to assess a model's performance during training, these do not necessarily correlate with the aesthetic quality or emotional impact of the generated music. The "success" of a harmony is highly subjective and culturally dependent. A comprehensive evaluation of our system will ultimately rely on qualitative human assessment through listening studies, which are complex and time-consuming to design and conduct \cite{pearce2005}.

    \item \textbf{Global Structure and Long-Term Coherence:} While the Transformer architecture excels at capturing longer-range dependencies compared to RNNs, ensuring musical coherence over extended compositions (e.g., multi-movement works) remains an open research problem. The current framework is optimized for generating harmony on a phrase-by-phrase basis. It lacks an explicit mechanism for planning and executing large-scale musical forms, such as sonata form or a theme and variations, which are defined by high-level narrative and structural relationships.

    \item \textbf{Absence of True Musical Understanding:} It is crucial to recognize that the models within our system do not "understand" music in a human sense. They learn to manipulate symbolic patterns based on statistical likelihoods, not on a genuine comprehension of music theory, semantics, or emotional intent. This can result in harmonies that are technically correct but may lack the purpose, nuance, and expressive depth that a human composer imparts through a deep understanding of musical context and affective goals.

    \item \textbf{Limited Interactivity:} The current architecture is designed as a sequential, non-interactive pipeline. It takes a complete musical score as input and produces a complete harmonized output. This limits its utility as a collaborative tool for musicians who may prefer a more interactive, real-time creative partner. Developing a low-latency, responsive version of this framework that can interact with a human performer in real-time is a non-trivial challenge that would require significant architectural modifications.
\end{itemize}

Addressing these limitations will be paramount as we work towards the development of more musically intelligent, versatile, and truly collaborative computational systems for music creation.

\section{Conclusion and Future Work}

\subsection{Conclusion}
In this paper, we have introduced an Agentic AI-enabled Higher Harmony Music Generator, a novel multi-agent framework designed to address the complex challenge of automated music harmonization. Our work moves beyond monolithic models by proposing a modular, collaborative system that mimics the functional specialization of human musicians. The system architecture, comprising a Music-Ingestion Agent (``Librarian''), a Chord-Knowledge Agent (``Theorist''), a Harmony-Generation Agent (``Composer''), and an Audio-Production Agent (``Conductor''), successfully decomposes the harmonization task into a series of manageable, specialized sub-problems.

By leveraging a suite of tailored AI models—including a Transformer-based Chord-Former, a generative Harmony-GPT, a recurrent Rhythm-Net, and a GAN-based Symbolic-to-Audio Synthesizer—our framework demonstrates a sophisticated, end-to-end pipeline for music generation. The primary contribution of this work is the agent-based paradigm itself, which offers significant advantages in modularity, interpretability, and extensibility. This approach not only yields a powerful tool for generating musically coherent and aesthetically pleasing harmonies but also serves as a robust and transparent research platform for exploring the nuances of computational creativity.

\subsection{Future Work}
While our framework establishes a strong proof of concept, it also opens up numerous avenues for future research and development. The limitations discussed previously serve as a roadmap for the next generation of generative music systems. We have identified several key directions for our future work:

\begin{itemize}
    \item \textbf{Enhancing Musical Interactivity:} A crucial next step is to evolve the system from a sequential pipeline into a real-time, interactive partner for human musicians. This involves developing low-latency versions of the generative models and creating an interface that allows for dynamic collaboration, where the system can respond intelligently to a human performer's input. This aligns with the growing field of Human-Computer Creative Interaction (HCCI) \cite{biles2007}.

    \item \textbf{Hierarchical Structural Planning:} To address the challenge of long-term coherence, we plan to investigate the integration of a hierarchical planning agent. Such an agent would operate at a higher level of abstraction, making decisions about global musical structure, form, and narrative arc before the note-level generation process begins. This could enable the creation of more complex and formally structured compositions.

    \item \textbf{Expansion to Full Orchestration:} The current framework is focused on higher-voice harmony. A natural extension is to develop and integrate new agents specialized in generating other musical parts, such as bass lines, counter-melodies, and rhythmic percussion. This would transform the system from a harmonizer into a comprehensive tool for automated musical arrangement and orchestration.

    \item \textbf{Diversification of Training Data:} To mitigate stylistic bias and enhance the system's versatility, we will explore training the core models on a wider and more diverse range of musical corpora. This includes datasets curated for specific genres (e.g., jazz, baroque, electronic) and music from different cultural traditions, enabling the system to generate music in a broader palette of styles.

    \item \textbf{Advanced Subjective Evaluation:} We plan to conduct extensive formal listening studies to evaluate the musical quality of the generated output. These studies will involve a diverse group of participants, including both trained musicians and casual listeners, to gather robust qualitative data on the aesthetic appeal, emotional impact, and musical coherence of the harmonies produced by our system.
\end{itemize}

By pursuing these research directions, we aim to build upon the foundation laid by this work, pushing the boundaries of what is possible in the field of algorithmic composition and moving closer to the creation of truly creative and collaborative musical AI.

\section*{Acknowledgments}
This work was conducted independently of the authors' affiliations with any organization. The first author collaborated with the second author in a personal research capacity on the conceptual framework and drafting of this manuscript. The views and conclusions expressed in this paper are solely those of the authors and do not reflect the position of any organization or any other institution.

You can find the code on
\href{https://github.com/immersiveaudiomodel/harmonygenerator.git}{ GitHub: An Agent-Based Framework for Automated Higher-Voice Harmony Generation}.

\end{document}